\documentclass[10pt,letterpaper]{article}
\usepackage{opex3}

\usepackage{amsmath}
\usepackage{amssymb}

\usepackage{graphicx}
\usepackage{float}

\usepackage{xspace}

\renewcommand{\deg}{^{\circ}}

\newcommand{\mcaobench}{test bench\xspace}
\newcommand{\PFCRMS}{\ensuremath{11.2\;\textrm{nm}}}
\newcommand{\PFCSR}{\ensuremath{0.520}}
\newcommand{\FTRRMS}{\ensuremath{35.0\;\textrm{nm}}}
\newcommand{\FTRSR}{\ensuremath{0.479}}
\newcommand{\ATMRMS}{\ensuremath{180.0\;\textrm{nm}}}
\newcommand{\WindFreqSix}{\ensuremath{-0.042\;\textrm{timesteps}^{-1}}}

\begin{document}
\title{A laboratory demonstration of an LQG technique for correcting frozen flow turbulence in adaptive optics systems}

\author{Alexander Rudy$^{1,*}$, Lisa Poyneer$^{2}$, Srikar Srinath$^{1}$, S. Mark Ammons$^{2}$, and Donald Gavel$^{3}$}

\address{$^1$University of California Santa Cruz, 1156 High Street, Santa Cruz, CA, USA\\
$^2$Lawrence Livermore National Laboratory, 7000 East Avenue, Livermore, California 94550, USA\\
$^3$University of California Observatories, 1156 High Street, Santa Cruz, CA, USA}

\email{$^*$arrudy@ucsc.edu}

\begin{abstract}
We present the laboratory verification of a method for removing the effects of frozen-flow atmospheric turbulence using a Linear Quadratic Gaussian (LQG) controller, also known as a Kalman Filter. 
This method, which we term ``Predictive Fourier Control,'' can identify correlated atmospheric motions due to layers of frozen flow turbulence, and can predictively remove the effects of these correlated motions in real-time. 
Our laboratory verification suggests a factor of $3$ improvement in the RMS residual wavefront error and a $10\%$ improvement in measured Strehl of the system. 
We found that the RMS residual wavefront error was suppressed from $\FTRRMS$ to $\PFCRMS$ due to the use of Predictive Fourier Control, and that the far field Strehl improved from $\FTRSR$ to $\PFCSR$.
\end{abstract}


\section{Introduction}
\label{sec:introduction}

We present progress towards the implementation of a Predictive Fourier Control (PFC) \cite{2007JOSAA..24.2645P,2008JOSAA..25.1486P} algorithm on sky to correct for frozen flow turbulence in adaptive optics systems. 
Recently, linear-quadratic Gaussian (LQG) controllers \cite{Gavel:2003ki,2004JOSAA..21.1261L} have been used for the stable control of adaptive optics systems \cite{2008OExpr..16...87P}, and for the suppression of specific frequency vibrations in operating adaptive optics systems \cite{2014SPIE.9148E..0KP,Beuzit:2008gt}. 
Predictive Fourier Control is an LQG method which explicitly corrects for frozen-flow turbulence crossing the telescope aperture. 
It diagonalizes that problem by working in Fourier space, making the method computationally feasible for high-order AO systems.

Several groups have formulated LQG controllers for the optimal control of adaptive optics systems. 
Gavel and Wiberg \cite{Gavel:2003ki} initially developed a Kalman filter for optimal AO control. 
Le Roux et al. \cite{2004JOSAA..21.1261L} formulated an LQG controller for adaptive optics. 
Petit et al. \cite{2008OExpr..16...87P} demonstrated the use of LQG in the laboratory to suppress vibrations. 
Sivo et al. \cite{2014OExpr..2223565S} have applied this on sky with the CANARY demonstrator.
LQG control has progressed from the laboratory to the latest generation of astronomical instruments.
The recently commissioned Gemini Planet Imager \cite{Macintosh:2014js} and SPHERE \cite{Beuzit:2008gt} instruments have both demonstrated the use of optimal LQG control \cite{2014SPIE.9148E..0KP,2014SPIE.9148E..0OP}.
GPI's AO system uses an LQG controller to suppress vibrations in both tip-tilt and focus \cite{2014SPIE.9148E..0KP}.
Similarly, SPHERE uses an LQG controller to suppress vibrations in only the tip-tilt modes \cite{2014SPIE.9148E..0OP}.
These controllers are implemented with the same LQG framework that we will apply to correct frozen flow turbulence across spatial frequencies beyond the tip-tilt regime.

Our experimental setup provides a tightly integrated way to test the use of an LQG controller both in the lab and on-sky.
The ShaneAO system with the ShARCs Camera on the Shane 3-m telescope at Lick Observatory \cite{2014SPIE.9148E..05G,Kupke:2012gm} provides an on-sky instrument. 
At the Lab for Adaptive Optics we have configured an adaptive optics \mcaobench \cite{Laag08} to mimic the hardware and physical conditions for ShaneAO.
These closely coupled systems allow us to test and develop techniques in a controlled lab before proceeding to the instrument on-sky for functional tests.

This paper describes a laboratory implementation of the full Predictive Fourier Control (PFC) algorithm.
In Section~\ref{sec:experimental_setup} we provide a brief overview of the \mcaobench, a full, physical simulator designed to test adaptive optics hardware, software and techniques. 
In Section~\ref{sec:theoretical_framework} we describe the PFC algorithm, and the way this algorithm interacts with an operating adaptive optics system. 
In Section~\ref{sec:experimental_results} we describe the implementation of the PFC algorithm in the lab, and the measured performance improvement.
Section~\ref{sec:discussion} discusses the significance of this laboratory demonstration.

Our framework, described in Section~\ref{sec:theoretical_framework} and diagramed in Figure~\ref{fig:block:loops}, consists of three components:
\begin{enumerate}
    \item Phase reconstruction as performed by a standard adaptive optics system.
    \item The Predictive Fourier Controller, which uses a Kalman Filter to provide a best estimate of the correction in the future.
    \item Fourier Wind Identification, which provides the state space model to the Predictive Fourier Controller.
\end{enumerate}

Fourier Wind Identification is described theoretically in Section~\ref{sub:fourier_wind_identification}, and demonstrated in the lab in Section~\ref{sub:wind_identification_on_the_mcao_bench}. 
We discuss the theoretical framework of the Predictive Fourier Controller in Section~\ref{sub:lqg_based_predictive_control}, and demonstrate its operation in Sections~\ref{sub:integrating_the_lqg_controller}~and~\ref{sub:applying_predictive_fourier_control_to_the_mcao_test_bench}.

\section{Experimental Setup}
\label{sec:experimental_setup}

\begin{figure}[t]
   \begin{center}
   \includegraphics[width=\textwidth]{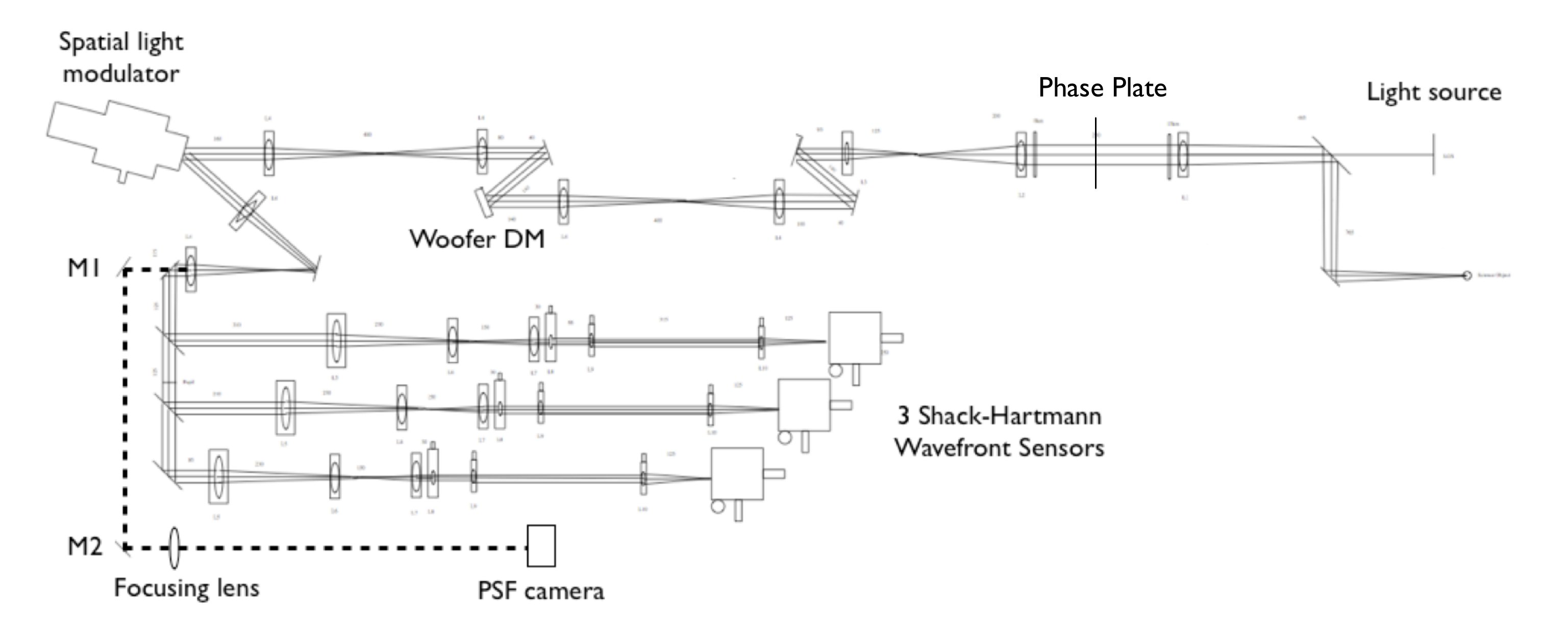}
   \end{center}
   \caption
   { \label{fig:testbed} 
  Testbed layout for wind prediction tests.
  The light source is a super-luminescent diode laser.
  The spatial light modulator is a high-frequency deformable mirror.
  In single conjugate mode only one of the Shack-Hartmann wavefront sensors is used.
  }
   \end{figure}

Our experiments are performed on a modified and optimized version of the UCSC Laboratory for Adaptive Optics (LAO) MCAO/MOAO testbed\cite{Laag08}. 
In the past, its utility and versatility has been shown by simulating MCAO and open-loop MOAO on a 30-m equivalent aperture telescope \cite{Ammons06}, demonstrating the positive effect of linearity calibrations to Shack-Hartmann wavefront sensors during open-loop MOAO operation on a 10-m telescope \cite{Ammons07} and for developing wavefront reconstruction and control algorithms for MCAO on an 8-m class telescope \cite{Laag08}. 
Currently, we have reconfigured the testbed to match as closely as possible the new adaptive optics system on the Shane 3-m telescope at Lick Observatory (ShaneAO) \cite{2014SPIE.9148E..05G} so that the transition from lab to a real system is as seamless as possible. The optical layout of the reconfigured system is shown in Figure~\ref{fig:testbed}.

For our purposes, the testbed is operated in single conjugate (SCAO) mode with two deformable mirrors (DMs): a low (spatial and temporal) frequency, high-stroke Alpao DM52 ``woofer'' and a high-frequency, lower-stroke, Hamamatsu SLM-X8267 spatial light modulator (SLM) ``tweeter''. As the \mcaobench does not operate in real time, we treat both mirrors as having instantaneous temporal properties and ignore the slower temporal performance of the ``woofer'' mirror.

Both DMs are conjugate to the equivalent of the atmosphere's ground layer represented by a phase plate mounted on a linear stepper motor.
Each phase plate is 160 subapertures across with a minimum wind speed of 0.01 subapertures per timestep.
The linear stepper can be reliably repositioned so that repeated runs can be performed across the same section of a phase plate and a higher signal-to-noise ratio is achieved by co-adding runs.
The size of the phase plate ensures that each test run contains $4.4$ pupil crossings.

The woofer's actuator layout is $4\times 4$ across the pupil.
The edges of the woofer are pinned, so outer actuators have reduced stroke and are not used.
This is nearly identical to the configuration of the woofer used in ShaneAO.
The SLM has a resolution of $768\times 768$, however pixels are binned to emulate a $36\times 36$ actuator configuration similar to that of the MEMS tweeter on ShaneAO.

Prior to every set of extended runs (typically once a day), the woofer and tweeter are re-registered to the wavefront sensor and a new woofer influence matrix is generated.
Woofer edge effects are minimized by performing a fit to individually measured influence functions.
A singular value decomposition (SVD) reconstructor zeroes out potentially problematic (high frequency) woofer modes.
Each DM has a separately set integrator leak gain (up to $0.99$ for the woofer and $0.995$ for the tweeter), similar to ShaneAO, to allow each mirror to handle its own actuator clipping and windup.
Detailed telemetry (residuals, DM commands) and configuration is saved during each run to ensure reproducibility.

The \mcaobench control law mimics the integrators and reconstructors used in ShaneAO, however the system delay can be arbitrarily shortened or lengthened to match the operating control delay measured with ShaneAO or to test the effects of longer or shorter time delays. 

\subsection{Modifications to the Experimental Setup to Support Predictive Fourier Control}
\label{sub:improving_the_mcao_bench}

In order to use the \mcaobench as a scale representation of the ShaneAO system, and to integrate the LQG controller into the \mcaobench software architecture, several changes to the \mcaobench were required. 

We first re-mapped the pupil of the test bench to the WFS lenslet arrays to simulate subaperutres which are scaled equivalents to the $10\;\mathrm{cm}$ subapertures found in the ShaneAO system.
\footnote{Throughout this paper, we use \mcaobench units, $\mathrm{timesteps}$ and $\mathrm{subapertures}$, as our test bench does not operate in real time.
All of these numbers can be easily scaled to match the $3\;\mathrm{m}$ ShaneAO system, an $8\;\mathrm{m}$ or a $30\;\mathrm{m}$ telescope. For the ShaneAO system, we convert from the \mcaobench to an operating frequency of $1\;\mathrm{kHz}$ and a subaperture size of $10\;\mathrm{cm}$.}
The remapping resulted in 36 active subapertures across the pupil, slightly more than are found on the ShaneAO system, which will have 30 subapertures across the pupil in its highest performance configuration.

As well, we implemented an SVD reconstructor for the system's high stroke woofer.
This allows us to eliminate problematic high frequency woofer modes, but to use the natural actuator basis set for the woofer directly.
This also allows us to easily re-measure the woofer influence functions for each actuator, recalibrating the system for minor changes in alignment and optimizing the system's long term performance.

Although the SVD reconstructor for the woofer improved its performance and registration, the SLM registration algorithm had to be improved to fix actuator misregistration errors that would cause very high spatial frequency content to build up on the mirror.
We solved this with an improved actuator registration method which uses actuators across the full SLM mirror working surface, and automatically fits the pixel locations instead of relying on user input to identify actuator pokes.

The original fiber laser was replaced with a super luminescent diode (SLD) to fix non-uniform pupil intensity effects that manifested as an $8\;\mathrm{Hz}$ power peak due to laser power fluctuations.
Over long runs (10,000 iterations or greater), we observe a correlation between RMS error and measured Strehl indicating low non-common path errors.
The far-field image is sharpened using an algorithm that iteratively maximizes Strehl by successively placing Zernike modes on the woofer and tweeter.

We also improved the control loop architecture of the \mcaobench to facilitate the use of the Kalman reconstructor.
The new architecture allows us to transition quickly between reconstructors, to perform calibration with one reconstructor (usually a simple Fourier-transform reconstructor) and then transition to a different reconstructor (the full LQG controller) for operation.
These modifications also permit the system to enable, disable and reset integrator states during normal operation.

\subsection{Error Budget for the Experimental Setup}
\label{sub:error_analysis_of_the_experimental_setup}

We have developed a consistent error budget for the \mcaobench in its current configuration, based on the error budget presented in Ammons et al. (2010) \cite{2010PASP..122..573A}. Similar to Ammons et al., we derive many of our error budget terms from the total atmospheric turbulence as measured by the wavefront sensor's RMS total wavefront error measured with the phase plate held still ($\sigma_{atmosphere}=\ATMRMS$). From this measurement, we can derive the fitting error ($\sigma_{fit} = 0.0441\sigma_{atmosphere}$, see \cite{2010PASP..122..573A} section 4.3.2) and aliasing error ($\sigma_{alias} = 0.4\sigma_{fit}$, see \cite{2010PASP..122..573A} section 4.3.2 and references therein). The measurement error from the wavefront sensor was measured directly on the \mcaobench, with no atmosphere present, and found to be $10.0\;\mathrm{nm}$.

We explicitly separate bandwidth and delay temporal errors. As the \mcaobench operates in discretized timesteps, there is no bandwidth error. The error due to the artificial delay in the system was computed from equation 9.56 in Hardy (1998) \cite{Hardy:1998},
\begin{equation}
    \sigma^{2}_\textrm{delay} = 28.4\left(f_{g} \tau_{s}\right)^{5/3}
\end{equation}
where $f_g$ is the Greenwood frequency, which can be computed from the wind velocity, and $\tau_{s}$ is the delay between measurement and correction, which for the experiments described in this paper was set to $\tau_{s}=2\;\mathrm{timesteps}$.

To measure the non-common path errors in our Strehl measurement, we applied the extended Mar{\'e}chal approximation (see Hardy (1998) equation 4.40 \cite{Hardy:1998}, and \cite{marechal1947etude}) to the far-field Strehl after image sharpening but with no aberrations in the system.
After image sharpening, the far-field Strehl was $0.80$, contributing $\sigma_{NCP}=49.5\;\mathrm{nm}$ of error to the system. Examining the error budget during experiments suggests that our far-field correction did not remain stable, and that an additional $44.0\;\mathrm{nm}$ of calibration error exist in the system. 
All of these error terms are presented in Table~\ref{tab:errorbudget}. 
The RMS far-field wavefront error in Table~\ref{tab:errorbudget} includes terms which are not seen by the wavefront sensor. 
When considering the RMS residual wavefront error as measured by the wavefront sensor, we ignore the Mirror fitting, spatial aliasing, static uncorrectable and calibration error terms. 
When measuring the far-field strehl, these terms are included.

\begin{table}
    \begin{center}
        \caption{\label{tab:errorbudget}Error Budget for the \mcaobench.}
    \begin{tabular}{l|r}
        Error Term & Value \\ \hline\hline
        Wavefront Sensing & $10.0\;\mathrm{nm}$ \\
        Mirror Fitting & $8.8\;\mathrm{nm}$ \\
        Spatial Aliasing & $3.5\;\mathrm{nm}$ \\
        Time Delay & $40.5\;\mathrm{nm}$ \\
        Static Uncorrectable & $49.5\;\mathrm{nm}$ \\
        Calibration & $44.0\;\mathrm{nm}$ \\ \hline
        RMS Far-Field Wavefront Error & $78.8\;\mathrm{nm}$ \\
        Predicted Strehl Ratio & $0.471$ \\
    \end{tabular}
    \end{center}
\end{table}

For the remainder of this paper, when we discuss RMS residual wavefront error, we are referring to only those terms which are seen by the wavefront sensor, or to measurements taken directly from the wavefront sensor. In contrast, the RMS total wavefront error is the wavefront error measured through our simulated atmosphere but with no correction applied. When considering the wavefront error in the far field, we will always convert to the Strehl ratio using the extended Mar{\'e}chal approximation.

\section{Theoretical Framework}
\label{sec:theoretical_framework}

We build on a Predictive Fourier Control (PFC) framework for Adaptive Optics Systems \cite{2007JOSAA..24.2645P,2008JOSAA..25.1486P} that aims to minimize temporal errors, including servo-lag, which manifest as a mis-estimation of the current atmospheric phase.
This predictive controller uses a slow ($\sim0.1\;\mathrm{Hz}$) loop to identify frozen flow turbulence seen by the AO system, and an LQG controller in the Fourier domain to suppress the specific errors which arise from frozen flow turbulence across the telescope aperture.

\begin{figure}[htbp]
  \begin{center}
    \includegraphics[width=\textwidth]{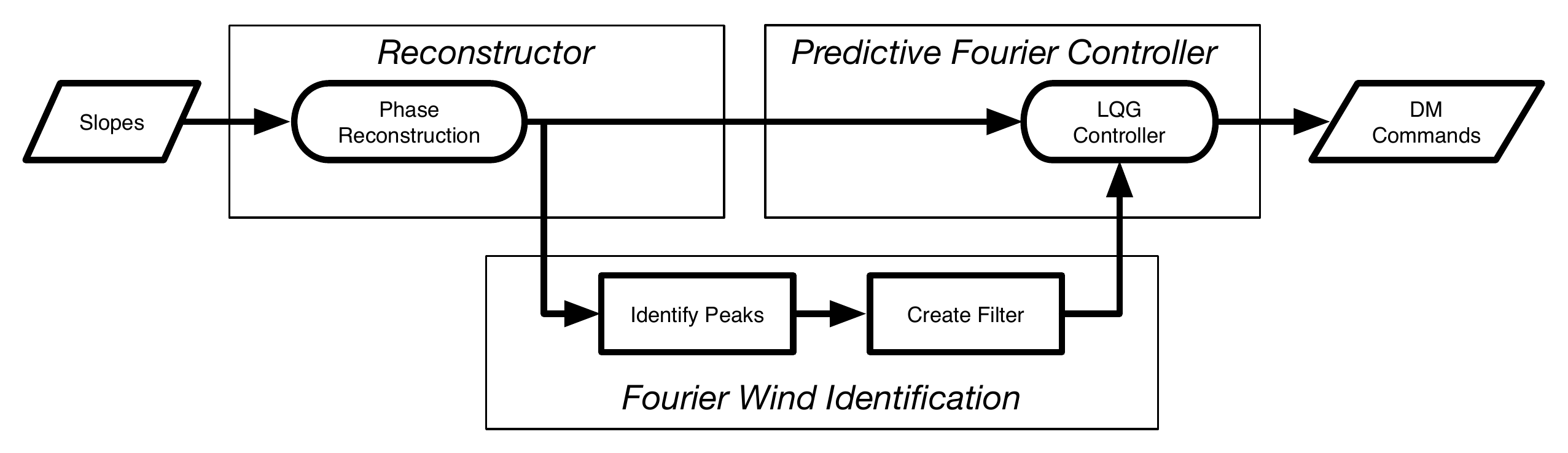}
    \caption{\label{fig:block:loops}
    A block diagram showing the dual-loop arrangement used with Predictive Fourier Control. The usual AO loop runs at $\mathrm{kHz}$ frequencies, while a separate analysis loop measures the frozen flow velocities and generates a new Kalman filter for the system. 
    Frozen flow turbulence is generally stable for more than 10s of seconds \cite{2009JOSAA..26..833P}, so updating the filter on these timescales should not affect the quality of the predictive control.
    However, even if the wind velocity does change before the filter is updated, we have shown (Section~\ref{sub:stable_lqg_control}) that the correction does not degrade significantly.
    }
  \end{center}
\end{figure}

The PFC algorithm uses telemetry from the adaptive optics system to identify frozen flow.
Telemetry is recorded over the course of $1$-$10$ seconds and then post-processed.
The post-processing uses the Fourier Wind Identification scheme (Section~\ref{sub:fourier_wind_identification}, and \cite{2009JOSAA..26..833P}), as well as a vibration and aliasing analysis to identify specific temporal frequencies which should be masked and eliminated. This information is used to construct a state space model that captures the dynamics of the atmosphere.
The relationship between the Predictive Fourier Controller and the Fourier Wind Identification is shown in Figure~\ref{fig:block:loops}.

This section describes the framework for Predictive Fourier Control, first describing Fourier Wind Identification (Section~\ref{sub:fourier_wind_identification}), then describing the Predictive Fourier Controller (Section~\ref{sub:lqg_based_predictive_control}).
The application of this framework to the \mcaobench is discussed in Section~\ref{sec:experimental_results}.

\subsection{Fourier Wind Identification}
\label{sub:fourier_wind_identification}

Using pseudo-open loop phase telemetry from the system, the Fourier Wind Identification (FWI) technique \cite{2007JOSAA..24.2645P} identifies atmospheric frozen flow for correction.
The Fourier basis set provides a convenient way to examine translating frozen-flow, as the individual Fourier modes each oscillate at a specific temporal frequency.

The temporal power spectral density (PSD) of each Fourier mode is estimated from the data.
Individual Fourier modes are split into segments of length $S$, which are windowed to emphasize the middle of the segment, and which overlap with neighboring segments. 
The overall frame rate, $f_s$ sets the maximum estimated temporal frequency at $f_s/2$. 
The length of the segments sets the frequency sampling spacing at $f_s/S$.
The half overlapped segments increase the signal-to-noise in the resulting PSD (for a more detailed discussion of this method, see \cite{2007JOSAA..24.2645P}).
By increasing or decreasing $S$, we can control the signal-to-noise and frequency resolution of the PSDs.

Once we have created a PSD for each Fourier mode, we can look for temporal peaks which are indicative of frozen flow.
In Fourier space, temporal peaks will appear with frequencies given by
\begin{equation}
  f_t = f_x v_x + f_y v_y\textrm{.} \label{eq:fwi:plane}
\end{equation}

\begin{figure}[htbp]
  \begin{center}
    \includegraphics[width=\textwidth]{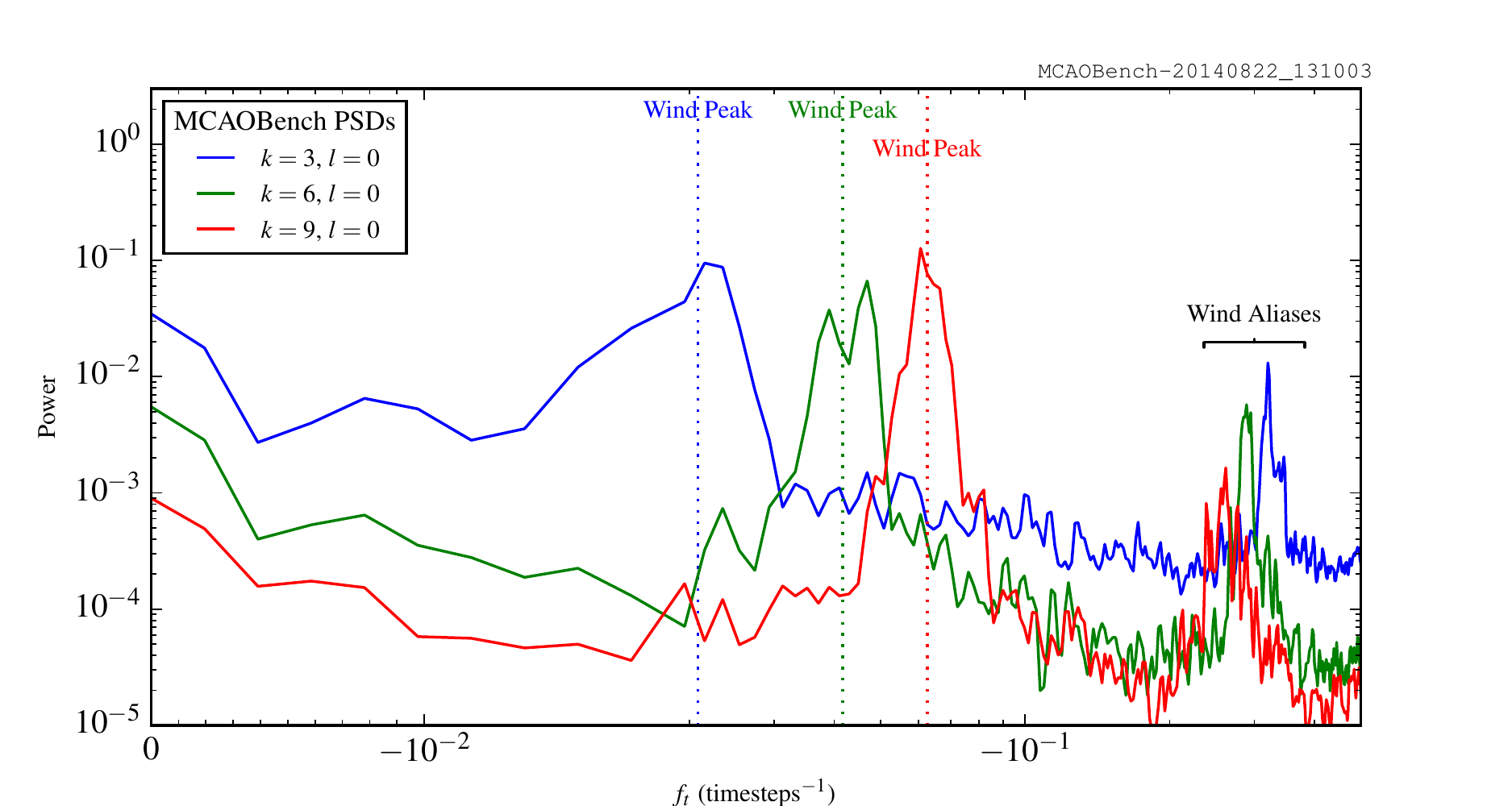}
    \caption{\label{fig:peaks}
    A temporal power spectral density (PSD) for a three separate Fourier modes ($k=3,6,9\;l=0$) which shows clear evidence of frozen flow turbulence.
    The system ran using an LQG controller which did not model any atmospheric frozen flow.
    Peaks at $\sim -0.021, -0.042,\;\textrm{and}\;-0.062\;\textrm{timesteps}^{-1}$ are due to the effect of frozen flow turbulence caused by a phase screen mounted on a stepper motor passing in front of the entrance pupil to the \mcaobench.
    The non-zero peak positions are set by Equation~\eqref{eq:fwi:plane}.
    The peaks at very high frequencies are spatial aliases of the frozen-flow wind.
    }
  \end{center}
\end{figure}

We model each peak as an oscillation at the temporal frequency $f_t$, with an added white noise broadening term \cite{2008JOSAA..25.1486P}.
Three sample PSDs from the \mcaobench, which show peaks that obey Equation~\eqref{eq:fwi:plane}, are shown in Figure~\ref{fig:peaks}.
Each peak corresponds to a potential match to Equation.
The peaks that appear close to $0\;\mathrm{timesteps}^{-1}$ are eliminated as they correspond to the slowly varying steady-state errors found in every system.
Peaks are fit for each spatial Fourier mode separately.

\begin{figure}[htbp]
  \begin{center}
      \includegraphics[width=0.49\textwidth]{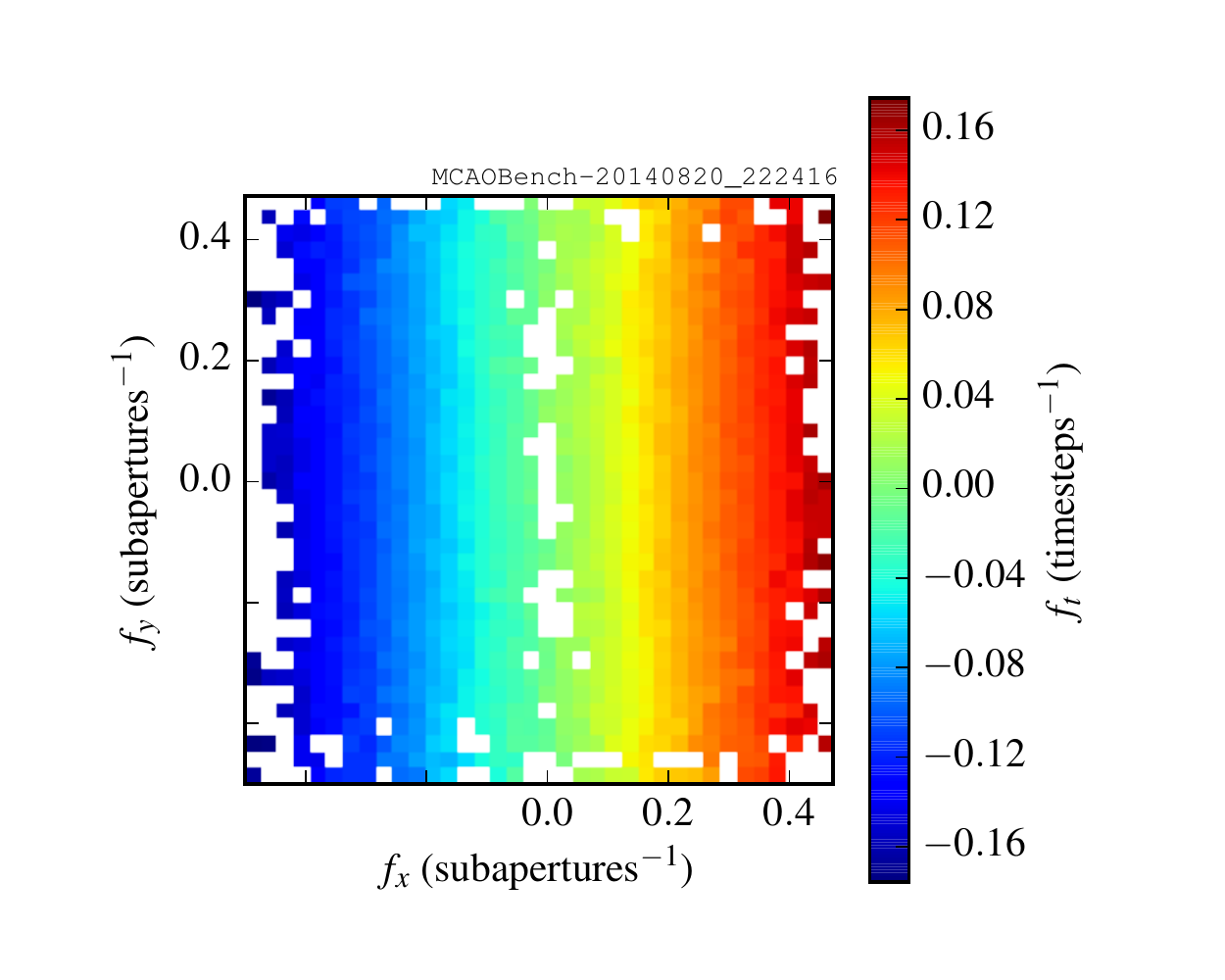}
      \includegraphics[width=0.49\textwidth]{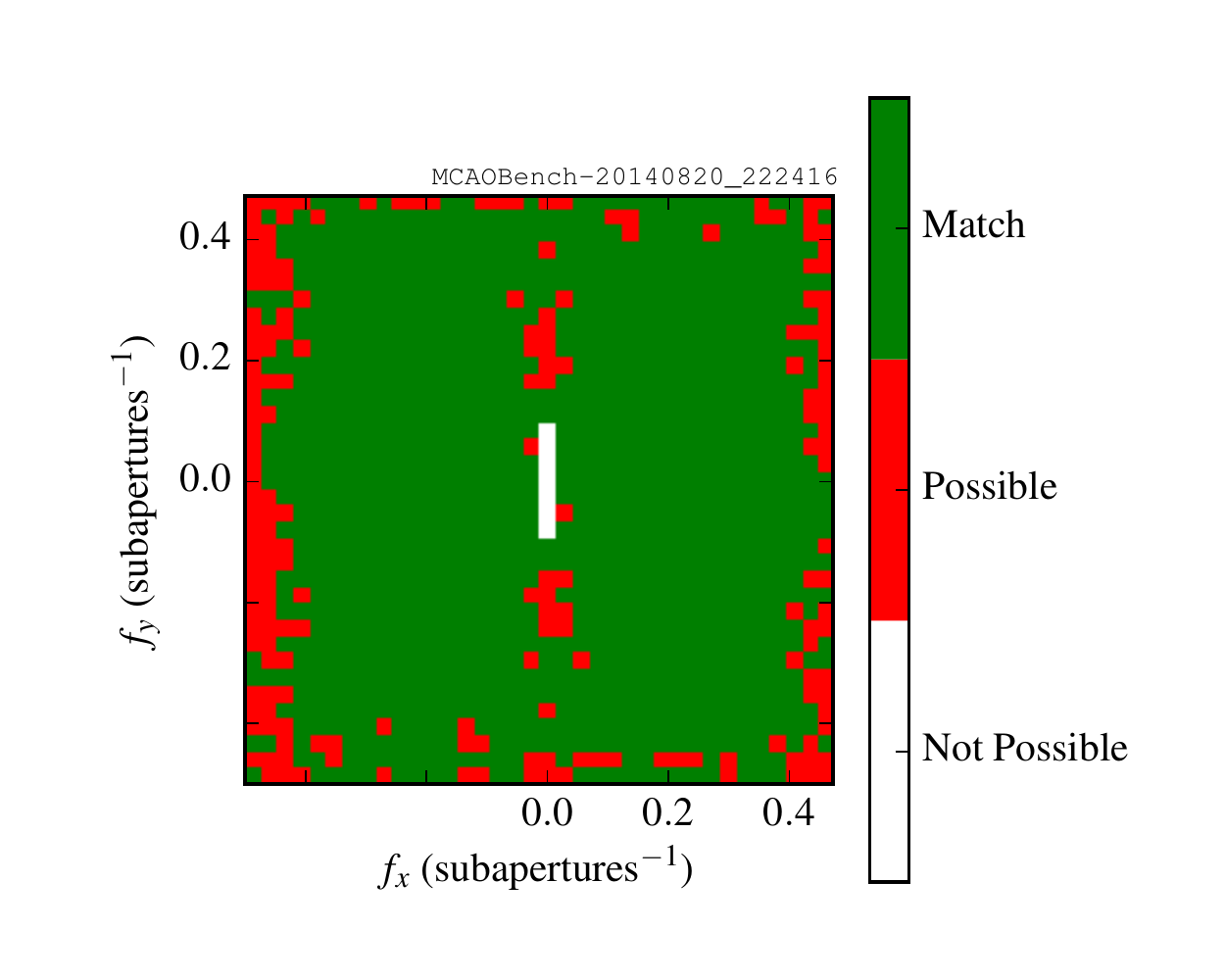}
    \caption{\label{fig:metric:matching}
    Identification of frozen flow turbulence on the \mcaobench using the Fourier Wind Identification algorithm. 
    This match shows a frozen flow layer that is moving in the $x$-direction ($v_x = 0.25$ and $v_y = 0.0$ $\mathrm{subapertures}/\mathrm{timestep}$).
    The left panel shows the detected peaks in telemetry. 
    The colorbar corresponds to the temporal frequency of the detected peaks. 
    Modes where no matching peak was found are shown as white.
    The right panel shows which peaks match the detected frozen flow layer. 
    It is red for modes which were not detected in the telemetry, and white for modes where the detected peak would be too close to $0\;\mathrm{timesteps}^{-1}$ to be detected. 
    For this set of telemetry data, $97\%$ of modes where peaks could be detected showed evidence of frozen flow. 
    }
  \end{center}
\end{figure}

Using the identified peaks from all the Fourier modes, the FWI algorithm works backwards through Equation~\eqref{eq:fwi:plane}. The frequencies in Equation~\eqref{eq:fwi:plane}, when shown on an $f_x$,$f_y$ grid, appear as a plane in frequency space, with the $f_x=0$,$f_y=0$ (piston) term always at $0\;\mathrm{timesteps}^{-1}$. Figure~\ref{fig:metric:matching} shows the process of matching found peaks in a PSD to a theoretical plane in Fourier space. FWI then produces a metric in velocity space that shows the percentage of matched peaks at each velocity. Areas with high metric scores are velocities at which frozen flow has been detected.

Once peaks have been identified, they can be formulated into a state-space model. 
Solving the Discrete Algebraic Riccati Equation (DARE) generates a controller that suppresses those specific temporal frequencies in its error transfer functions. 
The gains, leaks and high-pass coefficients of the LQG controller are then updated once the frozen flow layers and vibrations have been identified.

\subsection{LQG-Based Predictive Fourier Control}
\label{sub:lqg_based_predictive_control}

The wind parameters derived in the Fourier Wind Identification step are then used to calculate an optimal LQG controller for the AO system.
The controller uses a Kalman state-space framework with an independent state for each frozen flow layer.
A derivation of the appropriate Kalman filter for predictive control in a variable delay adaptive optics system is done analytically and numerically in \cite{2007JOSAA..24.2645P} and \cite{2008JOSAA..25.1486P}, and only a brief overview is provided below.

Frozen flow in the Fourier domain results in each Fourier mode shifting with a frequency given by Equation~\eqref{eq:fwi:plane}.
To apply this shift over a fixed delay between the phase measurement on the WFS and the application of the phase to the DM, we use a complex valued gain ($g_\alpha$ in the derivations from \cite{2007JOSAA..24.2645P}).
The complex gain serves to translate each Fourier mode to account for the delay between WFS sensing and command application, usually somewhere between 1 and 2 timesteps.
A purely real $\alpha$ corresponds to no frozen flow and a complex $\alpha$ determines the shift applied to each Fourier mode, with the complex phase\footnote{This should not be confused with the ``phase'' of the wavefront.} of $\alpha$ given by
\begin{equation}
  \mathrm{Phase}(\alpha) = - 2 \pi (k v_x + l v_y) \frac{T}{N d}
\end{equation}
where $k$ and $l$ are the respective Fourier mode numbers, $[v_x, v_y]\;(\mathrm{m}/\mathrm{s})$ is the wind vector for a single frozen flow layer, $T$ is the delay between the WFS measurement and the DM command application $(\mathrm{s})$, $N$ is the phase grid size ($N=36$ for the \mcaobench) and $d$ is the subaperture diameter $(\mathrm{m})$.
The magnitude of $\alpha$, $|\alpha|$, sets the ``memory'' for an individual layer, and is less than $1$.
Along with $\alpha$, the noise covariance matrix of the frozen flow layers (including the static term) is used to solve the DARE, and yield an optimal controller with $g_\alpha$ tuned for each Fourier mode.

The LQG controller can be applied directly with the Kalman Filter matrix equations. 
For our model, it can be simplified into the controller $C(z)$
\begin{equation}
  \label{eq:etf:pfc}
  C(z) = \left( Q^{-1} \sum_{k=0}^{L} \frac{p_{L+1,k} \alpha_k}{1-\alpha_k z^{-1}}\right) / \left( 1 - z^{-1} Q^{-1} \sum_{k=0}^{L} p_{L+1,k} \right)\textrm{,}
\end{equation}
which provides insight into how the prediction works. 
In Equation~\eqref{eq:etf:pfc} $Q = p + \sigma_v^2$ is the total layer power, and $\sigma_v^2$ is the noise variance of the particular Fourier mode.

This filter produces transfer functions which show a ``notch'' at the particular layer frequency, as well as good rejection around $0\;\mathrm{Hz}$, the DC term. 
The theoretical stability of these filters is examined in detail in \cite{2007JOSAA..24.2645P} and \cite{2008JOSAA..25.1486P}.

\section{Experimental Results}
\label{sec:experimental_results}

\subsection{Modeling Transfer Functions}
\label{sub:modeling_transfer_functions}

\begin{figure}[htbp]
  \begin{center}
    \includegraphics[width=\textwidth]{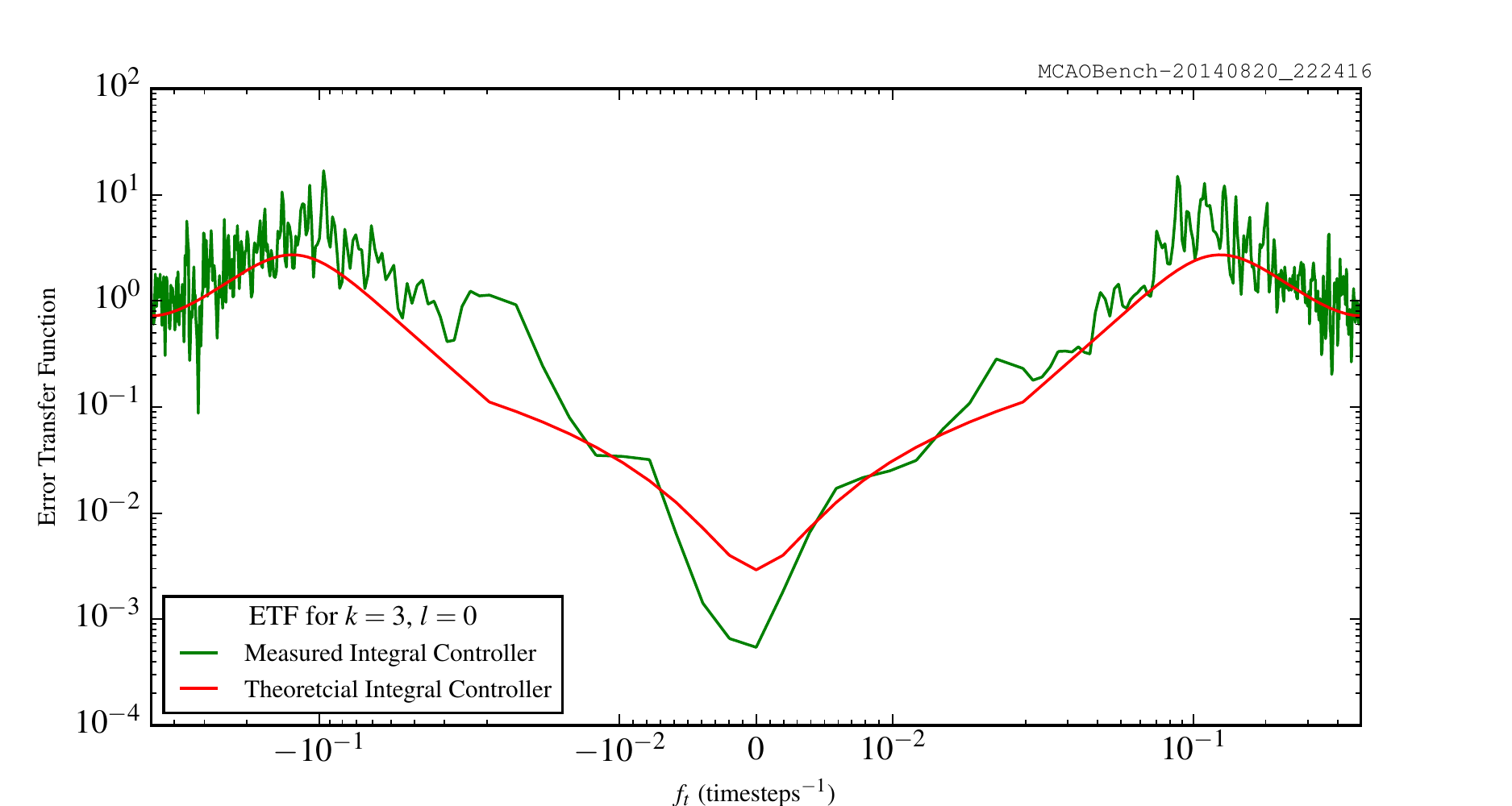}
    \caption{\label{fig:etf:good}
    An empirical (\emph{green}) and model (\emph{red}) error transfer function for the \mcaobench. 
    The empirical measurements were taken with an LQG controller which did not model any frozen-flow layers. 
    The model was computed using Equation~\eqref{eq:etf:pfc} for a 0-layer atmosphere (i.e. correcting for static errors only). 
    The transfer function shows the correct roll-off slope at low frequency. 
    The model also correctly predicts the controller bandwidth on the real system (the lowest frequency where the ETF first crosses $1$), though the model underestimates the overshoot (the degree to which the system makes frequencies higher than the bandwidth worse).
    }
  \end{center}
\end{figure}

With the improvements described in Section~\ref{sub:improving_the_mcao_bench}, we first demonstrate that the \mcaobench can reproduce theoretical transfer functions across a wide range of spatial frequencies for our system. 
To measure the error transfer functions (ETFs), we recorded the wavefront sensor residual error measurements for 4096 iterations with the control loop active and closed, and with the control loop uncontrolled and open.
For each telemetry data set, we converted to the Fourier basis, and then created periodograms using the technique described in Section~\ref{sub:fourier_wind_identification}. 
To compute the error transfer function, we took the ratio of the active, closed loop PSD for a given mode with the uncontrolled, loop open PSD. 
An ETF for a single Fourier mode ($k=5,l=5$) is shown in Figure~\ref{fig:etf:good} along with a model for that ETF.

Figure~\ref{fig:etf:good} shows the transfer function measured and computed for an LQG controller which does not model any frozen flow turbulence.
The LQG controller was implemented in a ``DC-only'' mode (simulating $0$ layers of atmospheric turbulence) which has similar performance characteristics to an integral controller used in many AO systems.

The model ETF was computed using Equation~\eqref{eq:etf:pfc}, but for a single, static layer, and so the model behaves like a standard integral controller. 
The improvements to the test bench allowed us to match the roll-off (the slope of the ETF at low frequencies), a feature which is universal to all first-order integral controllers, as well as the gain (which sets the left-right position of the model in Figure~\ref{fig:etf:good}) and bandwidth (the lowest frequency at which the ETF crosses $1.0$) between the model and the measured transfer function.

The well matched ETFs show that the system is performing nearly as expected at all spatial frequencies, and that more advanced controllers should be stable, with a predictable effect on the ETF of the system.

\subsection{Integrating the LQG Controller}
\label{sub:integrating_the_lqg_controller}

In order to support the LQG controller on our test bench, we made a few modifications to the control flow of the \mcaobench.
The LQG controller relies on an internal state vector and produces an estimate of the entire, open-loop phase at the next timestep.
However, both the \mcaobench, and the Shane-AO system use the phase error internally and apply the integrators only to the commands sent to the deformable mirrors.

In ShaneAO (and by default on the \mcaobench) the temporal integration is done on the mirror commands directly so that each integrator can consider and provide feedback for actuator clipping internally, without affecting the reconstructor or the state of the whole system. 
This is especially important when using multiple deformable mirrors to allow the system to take advantage of the high-stroke of the woofer and the many actuators on the tweeter.

To accommodate this architecture with LQG, we inserted a temporal differentiator into the phase reconstructor.
The differentiator is applied after the LQG controller is used to predict the phase at the next step.
Since the commands to each deformable mirror are independently integrated, the differentiation at the end of the LQG controller and the integration on each of the mirror commands are designed to cancel each other out.
This method (differentiating the open-loop correction, and applying to the woofer and tweeter separately) is used in GPI \cite{2014SPIE.9148E..0KP}.

However, as the individual mirror integrators for the woofer and tweeter DM have differing leak settings, and respond differently when actuators clip, the temporally differencing the output of the LQG controller is not guaranteed to produce a stable controller. 
When operating in this mode, the effective gain of the system is controlled by the coefficients of the LQG controller, while the individual command integrators are set to have a gain of $1$, but with their usual leak terms to prevent actuator saturation and windup.

\subsection{Demonstrating Stable LQG Control on the MCAO Test Bench}
\label{sub:stable_lqg_control}

Once the LQG controller was integrated into the \mcaobench, we were able to demonstrate simple stable LQG control.
The LQG controller achieved comparable RMS residual wavefront error ($\FTRRMS$ RMS residual wavefront error from $\ATMRMS$ RMS total wavefront error from the atmosphere) to the original integral controller over the course of 4,000 iterations.
The far field Strehl ratio remained comparably stable during this period.
The controller is stable for a variable frame delay of 1, 2, or 3 frames (higher delays were not tested).

Stability was easily maintained even when a wind suppression model was applied with no wind translation in the system.
The decrease in performance due to an incorrect frozen flow assumption was minimal, causing an RMS residual wavefront error of $50\;\mathrm{nm}$.

\subsection{Applying Fourier Wind Identification on the MCAO Test Bench}
\label{sub:wind_identification_on_the_mcao_bench}

We measured the Error Transfer Functions of the \mcaobench with a fast artificial wind speed ($0.25\;\mathrm{subapertures}/\mathrm{timestep}$). 
The system was run for 2048 iterations in both closed loop, and loop open (no correction) mode. Using the closed loop telemetry data, we were able to see evidence of frozen flow turbulence at every measured spatial frequency. 
Figure~\ref{fig:peaks} shows the temporal PSD for three Fourier modes ($k=3,6,9\;l=0$); each shows clear evidence for frozen flow turbulence.

Given the indication of frozen flow turbulence in the PSDs from the \mcaobench, we were able to construct a wind likelihood metric from test bench data showing that the method described in Section~\ref{sub:fourier_wind_identification} can successfully retrieve the input frozen flow wind velocity from the system telemetry. 
Figure~\ref{fig:metric:matching} shows the identification of the $0.25\;\mathrm{subapertures}/\mathrm{timestep}$ wind in \mcaobench telemetry.

The identified wind vector (or a-priori known wind vector(s)) can be provided to the filter generation algorithm described in Section~\ref{sub:lqg_based_predictive_control} and derived in \cite{2007JOSAA..24.2645P} and \cite{2008JOSAA..25.1486P} which will place a notch in the error transfer function for each Fourier mode to remove the temporal peak due to the frozen flow layer.
The ETF for $k=6,\;l=0$ will be notched near $\WindFreqSix$ and will reduce the large power peak visible in Figure~\ref{fig:peaks} for that mode. 

\subsection{Applying Predictive Fourier Control to the MCAO Test Bench}
\label{sub:applying_predictive_fourier_control_to_the_mcao_test_bench}

To test Predictive Fourier Control, we generated a state-space model for a known windspeed, assuming that the wind impacted all Fourier modes.
The filter was generated for wind speeds of $0.05$, $0.125$, and $0.25$ $\mathrm{subapertures}/\mathrm{timestep}$, and assumed a two frame delay to match our system design (see Section~\ref{sec:experimental_setup}).
To generate the filter, we assumed perfect frozen flow wind was detected in all Fourier modes.
This assumes that modes where the Fourier Wind Identification scheme (outlined in Section~\ref{sub:fourier_wind_identification}) did not detect peaks due to frozen flow turbulence are nevertheless worth including in the filter.
For the \mcaobench, this proved to be the correct assumption, but it should be tested on-sky as well.

\begin{figure}[htbp]
  \begin{center}
    \includegraphics[width=\textwidth]{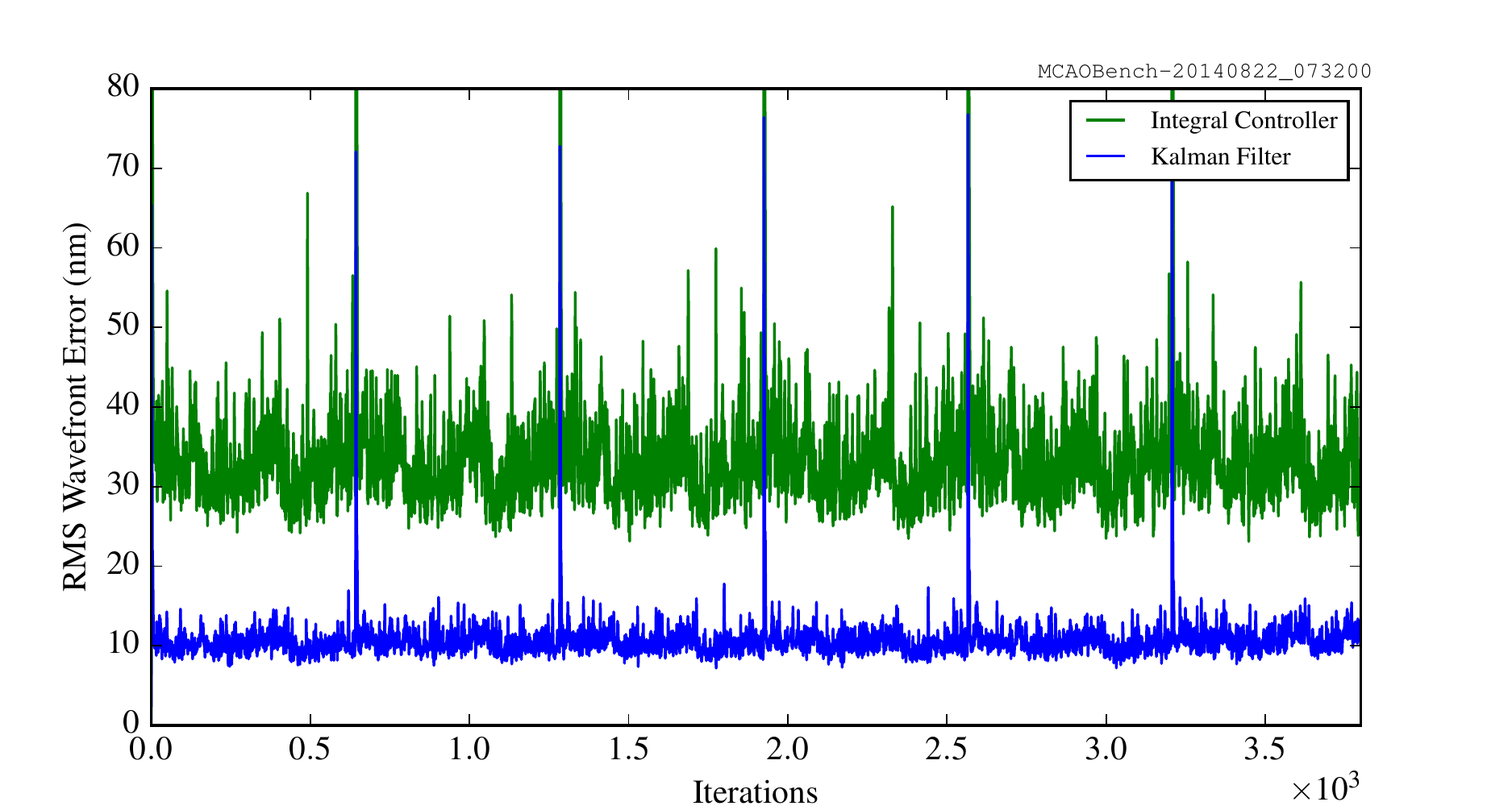}
    \caption{\label{fig:pkf:rms}
    The RMS residual wavefront error for both the Predictive Fourier Controller Kalman filter (\emph{blue}) and a standard integral controller (\emph{green}) over 4096 iterations. 
    For the standard integral, the RMS residual wavefront error was $\FTRRMS$, and for the Predictive Fourier Controller, it was $\PFCRMS$.
    The large spikes which occur every 640 iterations occur when the phase plate reaches the end of its movement range and must be reset to its initial position. 
    With no correction, the RMS total wavefront error is $\ATMRMS$.}
  \end{center}
\end{figure}

We compared the Predictive Fourier Control Kalman filter to a standard integral controller in the Fourier Domain, with a gain of $0.25$ and a leak of $0.99$.
The RMS residual wavefront error for both the Kalman Filter and the standard Integral Controller is shown in Figure~\ref{fig:pkf:rms}. 
The RMS total wavefront error (with no correction) is $\ATMRMS$.
The integral controller has a $\FTRRMS$ RMS residual wavefront error.
Predictive Fourier Control improves the RMS residual wavefront error by a factor of $3$ to $\PFCRMS$.

\begin{figure}[htbp]
  \begin{center}
    \includegraphics[width=\textwidth]{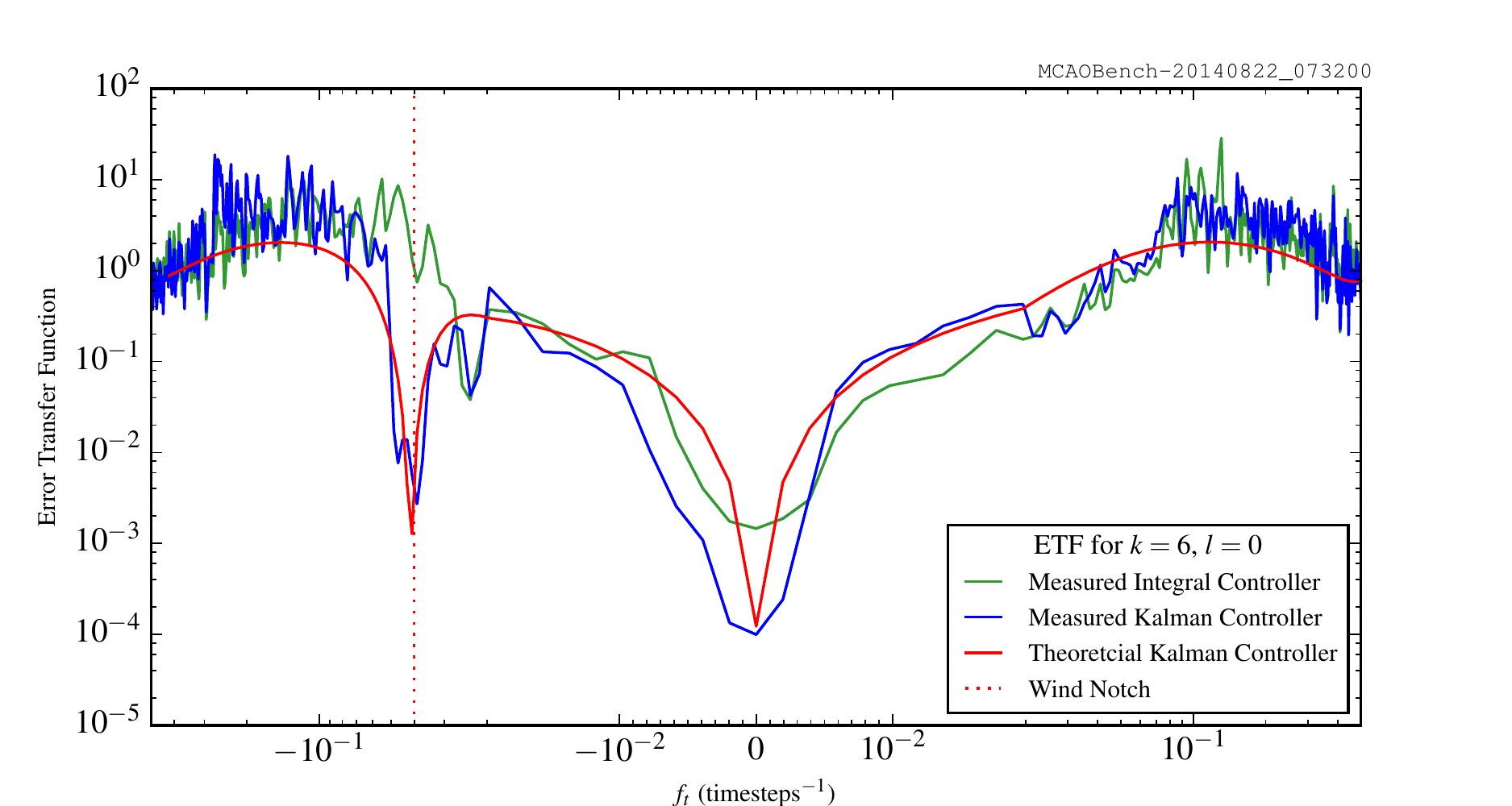}
    \caption{\label{fig:pkf:etf}
    An empirical (\emph{blue}) and model (\emph{red}) error transfer function for the \mcaobench using Predictive Fourier Control (PFC).
    For reference, we have plotted a measured transfer function using a standard integral controller (\emph{green}). The wind notch can be seen at $\WindFreqSix$.
    }
  \end{center}
\end{figure}

The measured Error Transfer Function of the \mcaobench when using Predictive Fourier Control is shown in Figure~\ref{fig:pkf:etf}.
The ETF was measured using the technique described in Section~\ref{sub:modeling_transfer_functions}.
The red dotted line in Figure~\ref{fig:pkf:etf} shows the expected temporal frequency of the wind for Fourier mode $k=6$, $l=0$ from Equation~\eqref{eq:fwi:plane}, $f_t = \WindFreqSix$. 
A clear ``notch'' in the PFC Kalman filter transfer function is visible at this frequency, where the error transfer function is reduced from $0.5$ to $10^{-3}$.
This behavior serves to remove the wind peaks visible in Figure~\ref{fig:peaks}, and so eliminates the dominant effect of frozen flow turbulence.

\begin{figure}[htbp]
  \begin{center}
    \includegraphics[width=\textwidth]{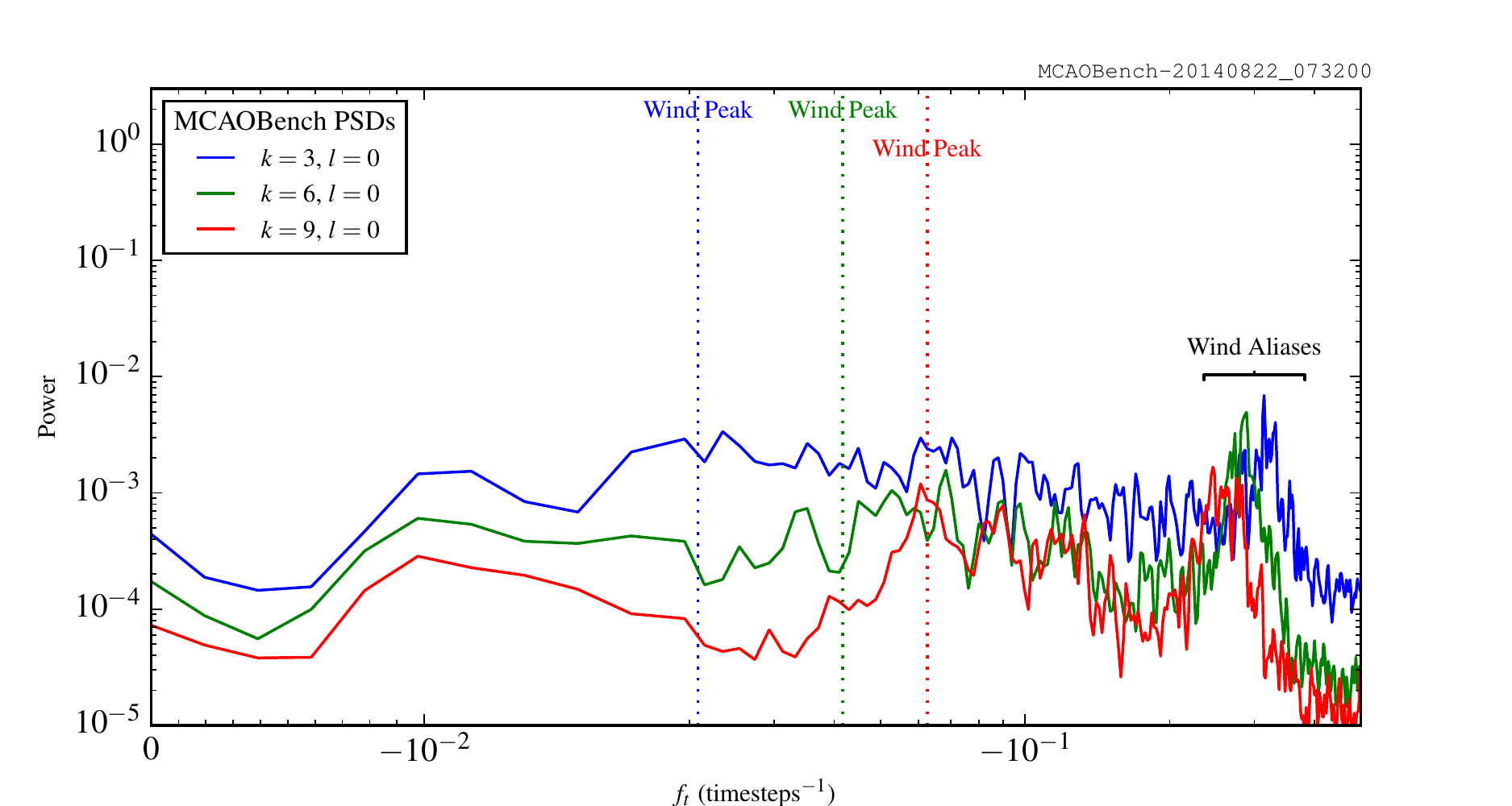}
    \caption{\label{fig:pkf:power}
    Same as Figure~\ref{fig:peaks}, but with the Predictive Fourier Control Kalman filter turned on.
    The frozen-flow peaks visible in Figure~\ref{fig:peaks} have been suppressed across all Fourier modes.
    The non-zero peak positions are set by Equation~\eqref{eq:fwi:plane}.
    The peaks at very high frequencies are spatial aliases of the frozen-flow wind.
    }
  \end{center}
\end{figure}

The elimination of these wind peaks can be seen in the periodogram generated from the PFC Kalman filter.
The power spectra for three Fourier modes ($k=3,6,9\;l=0$) are shown in Figure~\ref{fig:pkf:power}, which lacks the signature peaks from frozen flow that are visible in Figure~\ref{fig:peaks}.
The figure shows the power spectrum with the active closed loop.
Along with the expected suppression of power around $0\;\mathrm{timesteps}^{-1}$, there is clear evidence that the Predictive Fourier Control Kalman filter suppressed power at $\WindFreqSix$.
Additionally, correction can be applied at or beyond the controller bandwidth, as can be seen by the correction applied to $k=9$, $l=0$ in Figure~\ref{fig:pkf:power}.

\subsection{Far Field Performance Improvement}
\label{sub:far_field_performance_improvement}

\begin{figure}[htbp]
  \begin{center}
    \includegraphics[width=0.4\textwidth]{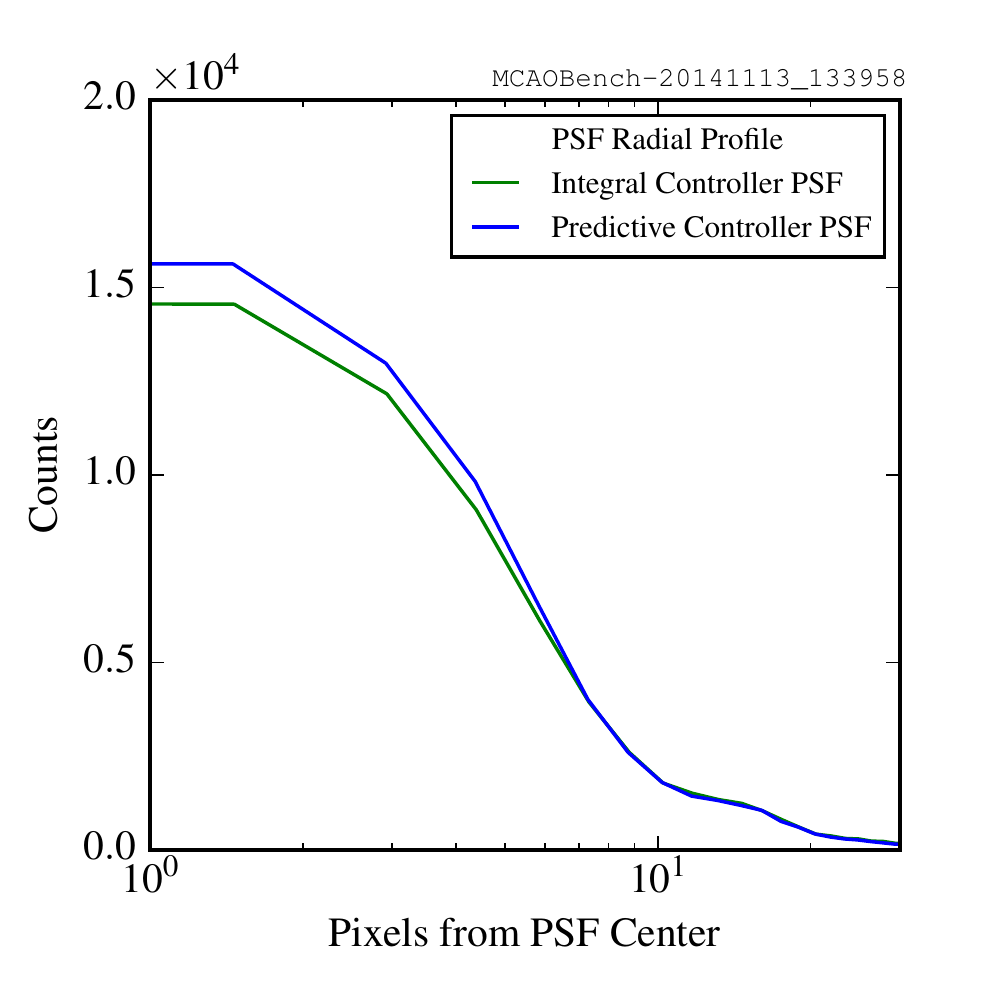}
    \caption{\label{fig:pkf:psf}
    Radial profile of the Point Spread Function of the \mcaobench showing the performance improvements between a standard integral controller (\emph{green}) and Predictive Fourier Control (\emph{blue}). 
    The Strehl improved from $\FTRSR$ to $\PFCSR$, a relative $10\%$ improvement in performance under otherwise identical conditions.
    }
  \end{center}
\end{figure}

To correctly predict the potential performance gains of Predictive Fourier Control on a working AO system, we also measured far-field images of our light source, which produced an aberrated and corrected point spread function (PSF).
The PSF was measured every 50 frames (due to the slower readout speed of the PSF camera) and then stacked to find the effective far-field image during closed-loop operation.
Using the integral controller, the \mcaobench produced a Strehl of $\FTRSR$.
With the predictive controller, the \mcaobench produced a Strehl of $\PFCSR$.
These measurements contain far-field images from multiple pupil crossings during each test run, and were reliably reproduced over multiple runs. 
Figure~\ref{fig:pkf:psf} shows the radial profile of the PSF for each case. 
Predictive Fourier Control improves the far-field Strehl performance by approximately $10\%$.
Gains on higher-Strehl systems should continue to be around $10\%$.

\subsection{Comparison of Error Terms}
\label{sub:comparison_of_error_terms}

We used the error budget described in Section~\ref{sub:error_analysis_of_the_experimental_setup} to compare the performance of the \mcaobench when using a standard integral controller, and our Predictive Fourier Control algorithm. A complete comparison of the error budgets for each controller is presented in Table~\ref{tab:errorterms}.

Using only the standard integral controller, we confirmed that the test bench can reproduce the expected RMS residual wavefront error and expected far-field performance. The \mcaobench over performed slightly (by $7\;\mathrm{nm}$) in RMS residual wavefront error, and underperformed in the far-field, which we took to be caused by additional non-common-path errors not captured by our image-sharpening routine, which, by inference from the performance of the standard integral controller, must be $44.0\;\mathrm{nm}$.

Then, comparing the performance of the test bench when using Predictive Fourier Control, we found the RMS residual wavefront error to be reduced by $80\%$ of the total time delay error. That the Predictive Fourier Controller did not remove $100\%$ of the time delay error is expected, as the Predictive Fourier Control transfer function cannot perfectly match the shape of the wind peaks in our simple model. This $80\%$ suppression is consistent between the residual wavefront measurements and the far-field Strehl measurements, reinforcing that Predictive Fourier Control should provide significant gains for on-sky adaptive optics systems.

\begin{table}
    \begin{center}
        \caption{\label{tab:errorterms}Comparison of Error Budget Terms between Predictive and Standard controllers.
        Terms that are not applicable to the wavefront sensor have been removed.
        From the final performance of the system, we estimate that about $20\%$ of the time delay error remains when using PFC.}
        \begin{tabular}{l|rrrr}
        \multicolumn{1}{c|}{Error Term} &  \multicolumn{4}{|c}{Experiment and Controller}  \\ \hline\hline
        Measurement Source & WFS & Far-Field & WFS & Far-Field \\
        Reconstructor & Standard & Standard & PFC & PFC \\ \hline
        Wavefront Sensing & $10.0\;\mathrm{nm}$ & $10.0\;\mathrm{nm}$ & $10.00\;\mathrm{nm}$ & $10.00\;\mathrm{nm}$ \\
        Mirror Fitting & $-$ & $8.8\;\mathrm{nm}$ & $-$ & $8.8\;\mathrm{nm}$ \\
        Spatial Aliasing & $-$ & $3.5\;\mathrm{nm}$ & $-$ & $3.5\;\mathrm{nm}$ \\
        Time Delay & $40.5\;\mathrm{nm}$ & $40.5\;\mathrm{nm}$ & $8.1\;\mathrm{nm}$ & $8.1\;\mathrm{nm}$ \\
        Static Uncorrectable & $-$ & $49.5\;\mathrm{nm}$ & $-$ & $49.5\;\mathrm{nm}$ \\
        Calibration & $-$ & $44.0\;\mathrm{nm}$ & $-$ & $44.0\;\mathrm{nm}$ \\ \hline
        RMS Wavefront Error & $41.7\;\mathrm{nm}$ & $78.8\;\mathrm{nm}$ & $12.9\;\mathrm{nm}$ & $68.1\;\mathrm{nm}$ \\
        Predicted Strehl Ratio & $-$ & $0.471$ & $-$ & $0.522$ \\ \hline
        Measured RMS Wavefront Error & $\FTRRMS$ & $77.1\;\mathrm{nm}$ & $\PFCRMS$ & $68.5\;\mathrm{nm}$ \\
        Measured Strehl Ratio & $-$ & $\FTRSR$ & $-$ & $\PFCSR$ \\
        \end{tabular}
    \end{center}
\end{table}

\subsection{Performance with Mismatched Filters}
\label{sub:performance_with_mismatched_filters}

To check the stability of the Predictive Fourier Controller, we ran the \mcaobench using filters generated for known incorrect wind speeds. 
The filters were generated in the same manner as in Section~\ref{sub:applying_predictive_fourier_control_to_the_mcao_test_bench}, but with the incorrect wind direction or magnitude for each test. 
Using a filter with an incorrect magnitude, or an incorrect direction up to $45\deg$ from the true wind direction resulted in stable correction, with performance degrading to that of the standard integral controller. 
Wind vectors that were significantly far from the true wind on the \mcaobench caused instabilities to appear in specific Fourier modes.

That the filters are stable for minor offsets of the wind vector is unsurprising. 
Filters with $\left|\alpha\right|<1$ should always be stable \cite{2008JOSAA..25.1486P}.
This suggests that the instabilities we observed with large wind-vector offsets are unstable not due to the controller alone, but due to the interaction of the controller and the \mcaobench.
Future experiments will investigate the specific modal instabilities found for large deviations in the wind vector.

\section{Discussion and Future Directions}
\label{sec:discussion}

This paper demonstrates the feasibility of ``Predictive Fourier Control'', but leaves a true characterization of system performance improvements and stability to future work.

The LQG filter was able to correct for frozen flow turbulence using our Fourier mode state-space model. Correction was applied in some modes beyond the controller bandwidth (see Figure~\ref{fig:pkf:etf}), and improved performance at very high temporal frequencies.

Although we have performed cursory checks of the LQG filter stability using the \mcaobench, several questions remain.

We plan to explore the stability of the LQG filter during real-world operation, where wind vectors can vary during the application of the filter. 
Although the filter appears stable for moderate changes in the wind vector, we will require an on-sky application to test the performance gains and losses that occur due to a naturally varying wind vector.

As the LQG filter maintains a state-space model of the system, it has some ``burn-in'' time during which it must build up a ``memory'' of the wavefront error.
We will characterize, on-sky, the effect of this ``burn-in'' time during filter transitions, and use this to optimize the trade-off between the speed of the Fourier Wind Identification loop (the bottom loop in Figure~\ref{fig:block:loops}) and the penalty incurred by changing the LQG filter in real time.

The PFC algorithm can also be extended to eliminate wind aliasing features, which appear as similar planes in Fourier space to those described by frozen flow.
In the same way that we can reduce delay error in the presence of frozen flow turbulence, we can use the LQG controller with the PFC algorithm to correct for the effects of wind aliasing when it is detected in system telemetry \cite{Veran:2010di}.

Finally, once we have demonstrated the PFC algorithm on-sky with ShaneAO, we intend to adapt the LQG controller to account for actuator clipping and the woofer-tweeter architecture.
A full controller would penalize actuators which are close to clipping within the Kalman filtering framework, and so would provide an end-to-end optimal control that would be less reliant on the absolute calibration of the individual woofer and tweeter mirrors.

\section{Conclusions}
\label{sec:conclusions}
This paper presents experimental verification of a computationally efficient method for removing the effects of frozen-flow turbulence from the residual errors in an adaptive optics system.
Predictive Fourier Control has the potential to reduce the effect of time delay errors, and to correct for specifically modeled and known errors beyond the bandwidth of the adaptive optics system.

We demonstrate that in the presence of wind-blown frozen flow turbulence Predictive Fourier Control is a feasible method for gaining $10\%$ improvement in an AO system in the far-field performance and a factor of $3$ in the RMS residual wavefront error.
The laboratory experiments suggest that we will see significant gains in AO system performance on-sky.
This level of performance improvement is promising for an algorithm which only requires a software adjustment, and which will not add significantly to the computational complexity of an existing adaptive optics system.

\section*{Acknowledgments}
\label{sec:acknowledgments}
This work performed under the auspices of the U.S. Department of Energy by Lawrence Livermore National Laboratory under Contract DE-AC52-07NA27344. The document number is LLNL-JRNL-667747. This work was funded by the UC Lab Fees Research Program grant 12-LF-236852. This material is based upon work supported by the National Science Foundation Graduate Research Fellowship Program under Grant No. DGE 1339067.

\end{document}